\documentclass[prl,twocolumn,superscriptaddress,preprintnumbers,%
amsmath,amssymb]{revtex4}

\usepackage{graphicx}

\makeatother

\newcommand{\be}{\begin{equation}}
\newcommand{\ee}{\end{equation}}
\newcommand{\bea}{\begin{eqnarray}}
\newcommand{\eea}{\end{eqnarray}}
\newcommand{\beann}{\begin{eqnarray*}}
\newcommand{\eeann}{\end{eqnarray*}}

\newcommand{\ba}{\begin{array}}
\newcommand{\ea}{\end{array}}
\newcommand{\bal}{\begin{align}}
\newcommand{\eal}{\end{align}}

\newcommand{\Tr}{\mathop{\rm Tr}}

\def\XXint#1#2#3{{\setbox0=\hbox{$#1{#2#3}{\int}$} 
\vcenter{\hbox{$#2#3$}}\kern-.5\wd0}}

\begin{document}


\title{M5-brane Defect and QHE \\ in  $AdS_4\times N(1,1)$/$\mathcal{N}=3$ SCFT}
 
\author{Mitsutoshi Fujita}
\affiliation{Department of Physics,
 Kyoto University, Kyoto 606-8502, Japan}


\begin{abstract}
We study the $d=11$ gravity dual $AdS_4\times N(1,1)$ of the $d=3$ $\mathcal{N}=3$ flavored Chern-Simons-matter (CSM) theory. In the dual gravity side, we analyze the M5-brane filling $AdS_3$ inside $AdS_4$ and derive the quantized Hall conductivity of the dual gauge theory. In the gauge theory side, this M5-brane intersects the gauge theory at the codimension-one defect.

\end{abstract}

\maketitle


\paragraph*{INTRODUCTION}---
$D=3$ $\mathcal{N}=6$ Chern-Simons-matter theory with the levels $(k,-k)$ for each node (the ABJM theory)~\cite{Aharony:2008ug} has recently been proposed as the effective theory of multiple M2-branes on the singularity of the orbifold $\mathbb{C}^4/\mathbb{Z}_k$. The susy-enhancement of the ABJM theory if $k=1,2$ was indicated in~\cite{Aharony:2008ug,Klebanov:2009sg} and for the general $U(N)\times U(N)$ gauge group, it was proved in~\cite{Samtleben:2010eu}.

The ABJM theory could also be explained from the IR limit of an elliptic brane setup in Type IIB string theory.  
Through T-duality and M-theory lift involving a 2-torus, the elliptic brane setup can be interpreted as the M2-branes transverse to a $d=8$ cone over $S^7/\mathbb{Z}_k$. After including the backreaction of $N$ M2-branes in the $d=11$ supergravity side, the gravity dual~\cite{Maldacena:1997re,Gub,Wit} beomes a solution of $d=11$ supergravity: $AdS_4\times S^7/\mathbb{Z}_k$. 
Generalizing their idea to yield elliptic 
${\mathcal N}=4$ SCFTs is explored in \cite{H}. 

Moreover, addition of the flavor degrees of freedom~\cite{Karch} to the ABJM theory was investigated in~\cite{Hohenegger:2009as, Gaiotto:2009tk, Hik,Amm, Jen} and for  $\mathcal{N}=4$ SCFT's, it was investigated in~\cite{Fuj2}.
 In the Type IIB string theory, adding massless flavors correspond to further placing D5-branes on the elliptic brane setup and preserve $\mathcal{N}=3$ superconformal symmetry. Through T-duality and M-theory lift, the D5-branes become the Kaluza-Klein monopole. The gravity dual of these SCFT's with flavors was proposed as the $d=11$ supergravity on $AdS_4\times \mathcal{M}_7$ where the cone over $\mathcal{M}_7$ is a $d=8$ toric hyperK{\"a}hler manifold~\cite{GGPT,Baba:2009nv} with special $sp(2)$ holonomy and with 3/16 supersymmetry. According to~\cite{Fuj2}, here,  $\mathcal{M}_7$ was proposed to be the Eschenburg space~\cite{Lee:2006ys} parametrized by three natural numbers $(t_1,t_2,t_3)=(pN_F,qN_F,kpq)$, where $p$, $q$, and $N_F$ correspond to the charges of NS5-branes, $(1,k)$5-branes, and D5-branes in the Type IIB elliptic brane setup, respectively.
In particular, the Eschenburg space with $t_1=t_2=t_3=1$ is interesting since it is equal to the regular tri-Sasaki manifold $N(1,1)$. The metric of $N(1,1)$ can be  easily written and the KK spectra of $N(1,1)$ was obtained in~\cite{Ter1,Fre1,Bil1}. However,  the dual $\mathcal{N}=3$ SCFT has been less known. 

The purpose of this paper is to progress the study of the AdS/CFT correspondence between the gravity dual $AdS_4\times N(1,1)$ and $\mathcal{N}=3$ SCFT. According to~\cite{Gaiotto:2009tk,Fuj2}, the flavored ABJM theory with the Chern-Simons levels $(1,-1)$ and a flavor is considered as the $\mathcal{N}=3$ SCFT dual to the $d=11$ supergravity on $AdS_4\times N(1,1)$ by analyzing its moduli space. 

We are interested in the application of our $AdS_4/\mathcal{N}=3$ SCFT correspondence to the 
Quantum Hall Effect (QHE)~\cite{Wen:2004ym} including the Fractional Quantum Hall Effect (FQHE)~\cite{Laughlin:1983fy} in the condensed matter physics. Here, the Hall conductivity in the QHE is quantized in units of $e^2/h$ and the longitudinal conductivity vanishes near the Plateau.  
There were holographic models describing the QHE~\cite{Kesk}-\cite{Bayntun:2010nx} and the recent holographic constructions of the FQHE~\cite{Fuj,Hik,Kawamoto:2009sn} inspired for M2-brane theories. 
 Moreover, non-abelian fractional Hall wavefunctions were obtained from M5 theory compactification~\cite{San} (see also \cite{AGT}).

 The benefits of the construction presented in this paper are the presence of the flavor field which is absent in the FQHE model of the ABJM theory~\cite{Fuj}. The flavor field is necessary for holographic duality and for the application to some realistic models in the condensed matter physics. Indeed, 
 the charge scalar field is needed in the Chern-Simons Ginzburg Landau description of the FQHE. Such a scalar field does not appear in the ABJM theory unless other fields such as flavors are added. 

In this paper, we consider the $AdS$/dCFT correspondence between the (1+1)-dimensional defect and the M5-brane filling $AdS_3$ inside $AdS_4$ instead of considering the bulk theory. 
Remind that the Hall current in the general QHE doesn't flow in the 2-dimensional space where electrons are localized but flow on an edge called the edge states~\cite{Wen:2004ym}.
According to~\cite{Fuj}, we introduce an edge M5-brane filling $AdS_3$ inside $AdS_4$ where the Hall current flows and derive the Hall conductivity of the dual flavored ABJM theory in the strong coupling region $N \gg k^5$ by analyzing the edge M5-brane.  {The M5-brane intersection in respect to the brane configurations of the 11-dimensional theory is also interesting since M2-branes corresponding to the $\mathcal{N}=3$ SCFT move the cone over the Eschenburg space or $N(1,1)$.} 

This paper is organized as follows: 1) We briefly review the $\mathcal{N}=3$ SCFT dual to $d=11$ supergravity on $AdS_4\times N(1,1)$.  2) We give the short review for the $\mathcal{N}=3$ $AdS_4$/SCFT correspondence. 3) We derive the QHE from the M5-brane worldvolume action in the dual gravity background. 


\paragraph*{$d=3$ $\mathcal{N}=3$ SCFT} --- 
{In the gauge theory side, we consider the $d=3$ ${\cal N}=3$ flavored ABJM theory with C-S levels $(1,-1)$ and a flavor proposed in the papers~\cite{Hohenegger:2009as, Gaiotto:2009tk, Hik}.
First, the ABJM theory without flavors consists of two gauge multiplets for the two copies of gauge group $U(N)_1\times U(N)_2$ and bi-fundamental chiral multiplets $(A_1,{A}_2)$ and $(\bar{B}_1,\bar{B}_2)$ in the $(\mathbf{N}_1,\bar{\mathbf{N}}_2)$ representation. The global symmetry of the ABJM theory at the classical level is baryonic $U(1)_b$ and $SU(4)_R$ $R$-symmetry.
 $D=3$ ${\cal N}=3$ flavored ABJM theory with C-S levels $(1,-1)$ and a flavor can be constructed by adding the D-term for the fundamental chiral-multiplets $(Q^1,\bar{\tilde{Q}}^1)$ transforming under the first gauge group as $\mathbf{N}_1$ to the ABJM theory and by modifying the superpotential.  $R$-symmetry is now broken to $SU(2)_R$, but the baryonic $U(1)_b$ stays unchanged, and the conformal symmetry is preserved.} 
 
The moduli space of $\mathcal{N}=3$ flavored ABJM theory which has C-S levels $(1,-1)$ and a flavor is discussed in~\cite{Gaiotto:2009tk}. This theory has the ring of chiral operators transforming as $\mathbf{8}$ under the flavor $SU(3)$ and so has the global $SU(3)\times SU(2)_R$ symmetry. That is, the isometry of these theory is the same as that of $N(1,1)$.
\paragraph*{$\mathcal{N}=3$ $AdS_4$/SCFT} --- 
We briefly review the dual $d=11$ supergravity on $AdS_4\times N(1,1)$. We 
 start with the Ricci-flat M-theory background $R^{1,2}\times \mathcal{M}_8$ without the backreaction of $N$ M2-branes. Here, $\mathcal{M}_8$ is the cone over the Eschenburg space $S(1,1,1)(=N(1,1))$.  
Since the transverse geometry is $d=8$ cone over $N(1,1)$, after the 
backreaction of $N$ M2-branes we have supersymmetry enhanced to a fraction 3/8 and are left with 
$AdS_4 \times N(1,1)$  
under 
the normalization 
\begin{eqnarray}
&ds^2_{11D} = \frac{R^2}{4} ds^2_{AdS_4} + R^2 ds_7^2 ,  \\
& N=\frac{1}{(2\pi \ell_p)^6}\int_{N(1,1)} \ast F_4,\ F_4=\frac{3}{8} R^3 vol_{AdS_4}, \\
&6R^6 vol(N(1,1))=\dfrac{3}{4}\pi^4R^6=(2\pi \ell_p)^6 N. 
\label{NOR}
\end{eqnarray}
Also, we have required $R_{ab} =6g_{ab}$ for the $N(1,1)$ metric. 
Note that $R=2R_{AdS}$ is the radius of 
$N(1,1)$. This background is regarded 
as the gravity dual of our ${\mathcal N}=3$ flavored ABJM theory. 

We write the metric of $AdS_4\times N(1,1)$~\cite{Gau,Pag} and the background flux $F_4$ as
\bea
&ds^2=\Big[\dfrac{dz^2+dx^{p}dx_{p}}{4z^2} +ds^2_{N(1,1)}\Big],\
F_4=6e^0e^1e^2e^3, \label{SUP21}
\eea  
where $e^{\bar{p}}$ $(\bar{p}=0,1,2,3)$ is the vierbein of the $AdS_4$ spacetime and we set the $N(1,1)$ radius 1.
Here, $ds^2_{N(1,1)}$ is the following metric on the manifold of an SO(3) bundle over $\mathbb{CP}^2$: 
\bea
&2ds^2_{N(1,1)} \nonumber \\
&=d\alpha^2+\dfrac{1}{4}\sin ^2\alpha (\sigma_1^2+\sigma_2^2)+\dfrac{1}{4}
\sin^2\alpha \cos^2\alpha\sigma_3^2 \nonumber \\
&+\dfrac{1}{2}\Big[ (\Sigma_1-\cos\alpha\sigma_1)^2+
(\Sigma_2-\cos\alpha\sigma_2)^2 \nonumber \\
&+(\Sigma_3-\dfrac{1}{2}(1+\cos^2\alpha)\sigma_3)^2 \Big], \label{MET42} \\
&\sigma_1=\sin\phi d\theta-\cos\phi\sin\theta d\psi ,\\
&\sigma_2=\cos\phi d\theta+\sin\phi\sin\theta d\psi, \ \sigma_3=d\phi +\cos\theta d\psi, 
\eea
with $0\le \alpha \le \pi/2$, $0<\theta<\pi$, $0<\phi<4\pi$ and $0<\psi<2\pi$. Here, $\Sigma_i$ are right-invariant 1-forms on SO(3), and $\sigma_i$ are right-invariant 1-forms on SU(2). According to~\cite{Pag}, $\Sigma_i$ must be of these forms in order for the part orthogonal to $\mathbb{CP}^2$ metric to be regular.   7-dimensional metric $ds^2_{N(1,1)}$ is scaled as $R_{mn}=6g_{mn}$. 
  
In $N(1,1)$, there are two important submanifolds $S^3$ at $\alpha=0,\ \Sigma_i=0$ and $S^3/\mathbb{Z}_2$ at $\alpha=0,\ \sigma_i=0$. Since both submanifolds are in $\alpha=0$ where the base $\mathbb{CP}^2$ metric vanishes, vierbeins of both submanifolds are also in the $SO(3)$ bundle direction. The position of $S^3$ and $S^3/\mathbb{Z}_2$ in $N(1,1)$ is investigated precisely in~\cite{Gau}.

Lastly, we want to comment on the M-circle. As stated in~\cite{Gaiotto:2009tk}, the position of the M-circle is the same as that in the GGPT geometry for the ABJM theory~\cite{GGPT}.
According to~\cite{Gaiotto:2009tk}, moreover, the dilaton in the IIA supergravity (or the coefficient of the M-circle) should not be constant in the internal manifold. Thus, the variables in $S^3/\mathbb{Z}_2$ should not be identified as the M-circle and so we regard the M-circle as an angular variable in $S^3$~\footnote{It will be difficult directly to compare the $d=8$ GGPT geometry with the $d=7$ $N(1,1)$ by making the cone structure clear. I would like to thank T. S. Tai for the comments on this point.}.
When we reduce the M-theory along this M-circle, then, we obtain the Type IIA string theory with dilaton depending on the $d=6$ internal manifold and leave $S^3/\mathbb{Z}_2$ a submanifold which the probe D6-brane can wrap. This submanifold is very similar to the lens space $S^3/\mathbb{Z}_2$ inside $\mathbb{CP}^3$ which appeared in the Type IIA gravity dual $AdS_4\times \mathbb{CP}^3$ of the ABJM theory. Then, the probe D6-brane  can also wrap this lens space in the gravity dual of the ABJM theory.

\paragraph*{QHE from the M5-brane worldvolume action}---
In this section, we analyze the M5-action~\cite{PST,Aga,Aga2} on $AdS_3\times S^3$ and derive the Quantum Hall Effect (QHE).  Since we use the dimensional reduction, our analysis is valid in the low energy limit and a part of $d=11$ metric reduces to an 1-form as seen in the reduction to the IIA supergravity. We also break the gauge symmetry $U(N)\times U(N)$
 to the $N$ copies of $U(1)\times U(1)$ since the QHE will be described by the $U(1)$ Chern-Simons theory. 

The M5-brane is wrapped on the submanifold $AdS_3\times S^3$ parametrized by  $\alpha=\Sigma_1=\Sigma_2=\Sigma_3=0$.  The 6-dimensional worldvolume coordinates are parametrized by $\xi^{\hat m}$ $(\hat m=0,...,5)$: $\xi^0=t,\ \xi^1=x^1,\ \xi^2=z,\ \xi^3=\psi,\ \xi^4=\theta, \ \xi^5=\phi'=\phi/2$.
The induced metric and the gauge field $a$ on the M5-brane are given by
\bea
&ds^2=\dfrac{1}{4}ds^2_{AdS_3} \nonumber \\
&+\dfrac{1}{4}\Big(d\theta^2 +\sin^2\theta d\psi^{ 2}) +\Big(d\phi'+\dfrac{1}{2}\cos\theta d\psi\Big)^2, \nonumber \\
&a=\dfrac{1}{2}\cos\theta d\psi ,
\eea
where $\int_{S^2} F'=2\pi$ $(F'=da)$. 

The 6-dimensional metric $G_{\hat m\hat n}$ contains 5-dimensional pieces $G_{mn},G_{m 5},G_{55}$. By setting $G_{5,5}=1$ and $\xi^5=\phi'(=X^{11})$, we can represent $G_{\hat m\hat n}$ by using the $6\times 6$ matrix
\bea
G_{\hat m\hat n}=\begin{pmatrix}
& G_{mn}^{(6)}+a_{m}a_{n} & a \\
& a^{\top }  & 1 \\
\end{pmatrix},
\eea
where we have rewritten $G_{5m}$ as $a_{m}$ for convenience since after the dimensional reduction, $a_{m}$ become 1-forms. 
We introduce the self-dual tensor gauge field by $B_{mn}$ and $H_{mnr}=3\partial_{[m}B_{nr]}$. Dual of $H_3$ becomes $\tilde{H}^{mn}=\epsilon^{mnrls}H_{rls}/6$, where $\epsilon^{mnrls}$ is the 5-dimensional flat epsilon symbol.
It is convenient to define $\mathcal{H}_{3}=H _{3}-b_{3}$,
where $b_{3}$ is the 3-form in the 11-dimensional supergravity.

The action of the M5-brane~\cite{Aga,Aga2} is $L_1+L_2$ where
\bea
&L_1=-\sqrt{-G}\sqrt{1+z_1+\dfrac{1}{2}z_1^2-z_2}, \\
&L_2=\dfrac{1}{8}\epsilon_{mnrls}\dfrac{G^{5r}}{G^{55}}\tilde{H}^{mn}\tilde{H}^{ls}, \\
&z_1=\dfrac{G_{mn}\tilde{H}^{nr}G_{rl}\tilde{H}^{lm}}{2(-G_5)}\equiv \dfrac{\Tr (G\tilde{H}G\tilde{H})}{2(-G_5)}, \\
&z_2=\dfrac{\Tr (G\tilde{H}G\tilde{H}G\tilde{H}G\tilde{H})}{4(-G_5)^2}.
\eea
Here, $G$ is the 6-dimensional determinant, $G_5$ is the 5-dimensional determinant written with $G_{mn}+a_{m}a_{n}$, and $G^{\hat m\hat n}$ is the inverse of $G_{\hat m\hat n}$. We have not written WZ term, since we don't need it for later analysis.

By dropping all dependence on $\xi^5$, we obtain the following 5-dimensional action:
\bea
&S=-\dfrac{1}{(2\pi)^4
}\int d^5\xi\Big(\sqrt{-G}\sqrt{1+z_1+\dfrac{z_1^2}{2}-z_2} \nonumber \\
&+\dfrac{\epsilon_{mnlst}a^{m}\tilde{H}^{nl}\tilde{H}^{st}}{8(1+a^2)} \Big).
\eea 
 
This action is equal to the following action~\cite{Aga2} with the Lagrange multiplier
\bea
&S_{5d}=-\dfrac{1}{(2\pi)^4}\int d^5\xi \Big(\sqrt{-\det (G_{mn}+2\pi\mathcal{F}_{mn})}+ \nonumber \\ &\int\Big(2\pi\mathcal{H}\wedge \mathcal{F}-\dfrac{(2\pi)^2}{2}a\wedge \mathcal{F}\wedge \mathcal{F}\Big)\Big), \label{M5A015}
\eea
where $\mathcal{F}_2$ is the worldvolume 2-form field strength assumed to be the 2-form on $AdS_3$. 
Note that the EOM of $\mathcal{F}$ gives $\tilde{\mathcal{H}}$ as the following function of $\mathcal{F}$ and $a$:
\bea
&\tilde{\mathcal{H}}=\dfrac{2\pi}{2}\sqrt{-\det (G_{\mu\nu}+2\pi\mathcal{F}_{\mu\nu})}\mathcal{F}^{\mu\nu}+2\pi \hat *a\wedge \mathcal{F}, \label{HTI016}
\eea
where $\hat *$ is the 5-dimensional flat epsilon symbol.

By integration by parts and by integrating $F'=da$, the 5-dimensional action \eqref{M5A015} reduces to the DBI+Chern-Simons action
\bea
&S_{5d}=-\dfrac{1}{(2\pi)^4}\int d^5\xi \sqrt{-\det (G_{mn}+2\pi\mathcal{F}_{mn})} \nonumber \\
&+\dfrac{1}{4\pi}\int A\wedge F \nonumber \\ 
&=-\dfrac{1}{(4\pi)^4} \int dx^0 dx^1 dz
\dfrac{1}{z^3}\cdot \nonumber \\
&\cdot\sqrt{1+32\pi^2 (z^4F_{x^1z}^2-z^4F_{x^0z}^2-z^4F_{x^0x^1}^2)} \nonumber \\
&+\dfrac{1}{4\pi}\int A\wedge dA. \label{M5d315}
\eea
 We shortly review how to derive the Hall conductivity from \eqref{M5d315}. We add
 the following boundary term~\cite{Eli} at $z=\epsilon$:
 \bea
 S_{bdy}=\dfrac{1}{4\pi}\int_{z=\epsilon}dx^0dx^1A_{x^0}A_{x^1}.
 \eea
 If the condition $\partial_{x^0}A_{x^1}=0$ is satisfied, then, the full action is invariant under the gauge transformation $\delta A=d\chi$ as follows:
\bea
&\delta \dfrac{1}{4\pi}\Big(\int A\wedge dA +\int dx^0dx^1A_{x^0}A_{x^1}\Big) \nonumber \\
&=\dfrac{1}{2\pi}\int dx^0dx^1\delta A_{x^0}A_{x^1} \nonumber \\
&=-\dfrac{1}{2\pi}\int dx^0dx^1 \chi \partial_{x^0}A_{x^1}=0.
\eea
 This condition means
  that we can still fix the worldvolume electric flux $F_{x^0x^1}\neq 0$. Remind also that the on-shell variation of the Chern-Simons term and the boundary term have only one of the two variables $\delta A_{x^0},\delta A_{x^1}$: $A_{x^0}$ and $A_{x^1}$ are considered as a pair of canonical variables with respect to $z$. To compute the Hall conductivity, moreover, the following boundary conditions should be satisfied:
\bea 
\delta A_{x^0}|_{bdy}=0 \quad \text{and} \quad \delta A_{x^1}|_{bdy} \quad \text{free}. \label{free20}
 \eea 
 The Hall conductivity can be derived from the continuity equation
\bea
J_{x^2}=-\left(\dfrac{\partial \rho}{\partial x^0}+\dfrac{\partial j_{x^1}}{\partial x^1}\right). \label{CON11}
\eea 
To compute $j_{x^1}$ and $\rho$ holographically, we need the EOM of \eqref{M5d315} and the relation given in~\cite{Gub,Wit}.
   
The equation of motion for \eqref{M5d315} are
\begin{eqnarray} &
\partial_{x^1}\left(\frac{\alpha zF_{x^1z}}{\sqrt{D}}\right)-\partial_{x^0}\left(\frac{\alpha zF_{x^0z}}{\sqrt{D}}\right)+\frac{1}{2\pi}F_{x^0x^1}=0,\nonumber \\ & \partial_z \left(\frac{\alpha zF_{x^1z}}{\sqrt{D}}\right)+\partial_{x^0}\left(\frac{\alpha zF_{x^0x^1}}{\sqrt{D}}\right)+\frac{1}{2\pi}F_{x^0z}=0,\\ & \partial_z\left(\frac{\alpha zF_{x^0z}}{\sqrt{D}}\right)+\partial_{x^1}\left(\frac{\alpha
zF_{x^0x^1}}{\sqrt{D}}\right)+\frac{1}{2\pi}F_{x^1z}=0,
\label{eomym}
\end{eqnarray} where $D=1+32\pi^2(z^4F_{x^1z}^2-z^4F_{x^0z}^2-z^4F_{x^0x^1}^2)$ and $\alpha$ is a constant.
The boundary currents are given by
\bea
&\rho =\dfrac{\delta S}{\delta A_{x^0}}\Big|_{z=0}=-\dfrac{\alpha zF_{x^0z}}{\sqrt{D}}+\dfrac{1}{2\pi}A_{x^1}\Big|_{z=0},\\
& j_{x^1}=\dfrac{\delta S}{\delta A_{x^1}}\Big|_{z=0}=\dfrac{\alpha zF_{x^1z}}{\sqrt{D}}\Big|_{z=0}, \label{CUR319}
\eea
where we fixed the potential $A_{x^0}|_{z=0}$ after the functional differentiation. This condition is consistent with \eqref{free20} (see also~\cite{Jensen:2010em}). 
By substituting \eqref{CUR319} and \eqref{eomym} into \eqref{CON11}, we obtain the Hall conductivity:
\bea
&j_{x^2}=-\dfrac{1}{2\pi}\partial_{x^1}A_{x^0}=\dfrac{1}{2\pi}E_{x^1}, \\
&\to \sigma_{x^1x^2}=\dfrac{e^2}{h},\ \nu =1, \label{HAL31}
\eea
where we recovered the Planck unit $\hbar =h/2\pi$ and the electric charge $e$.
Here, $\nu=1$ is the filling fraction. Thus, the Hall conductivity is quantized in terms of the C-S level of the dual theory $k=1$ and there are no other contributions (see also~\cite{Hik}). 

 It will also be interesting to derive conductivities of the boundary liquid in particular in the finite temperature~\cite{Hun}. However, the full EOM of \eqref{M5d315} seems not to be solved analytically and the finite temperature GKPW relation of the DBI-CS theories on $AdS_3$ seems to be complicated.
 
\paragraph*{The stability of the M5-brane embedding} --- 
It will be important to investigate the stability of the M5-brane embedding, computing the fluctuations from the $AdS_3\times S^3$ induced metric. 
Here, $S^3$ is a trivial-cycle~\cite{Karch2,Karch} on $N(1,1)$. For this purpose, it is convenient to use the PST action~\cite{PST} for the M5-brane. To investigate the fluctuation of the PST action, the DBI part is only needed~\cite{Arean:2006pk} if we assume no fluxes along $S^3$ and impose a gauge such that the auxiliary PST scalar $a$ equals the coordinate $x^1$. Here, the fluctuation $\alpha =0+\bar{\alpha}$ is considered. Expanding the DBI action to quadratic order in the fluctuations $\bar{\alpha}$ and choosing the zero mode of $S^3$ harmonics, the follwing action is obtained:
\begin{align}
&\mathcal{L}_{DBI}= \sqrt{-G}\Big( 1+R^2_{AdS}G^{ab}\partial_{a}\bar{\alpha}\partial_{b}\bar{\alpha}-\dfrac{3}{4}\bar{\alpha}^2\Big),
\end{align}
where we recovered the AdS radius $R_{AdS}=R/2$ and $a,b=0,1,2$.
Then, the square of mass $-3/(4R_{AdS}^2)$ for $\bar{\alpha}$ is above the BF-bound $m^2_{BF}=-1/R_{AdS}^2$~\cite{Breitenlohner}. Thus, our M5-brane is stable~\footnote{Another M5-brane describing different physics is the M5-brane wrapped on $S^3/\mathbb{Z}_2$ in $N(1,1)$ different from $S^3$. This embedding can easily be shown to be stable along the direction $\alpha$.}.

The situation is similar to the M5-brane embedded in $AdS_4\times S^7$. 
 Indeed, the M5-brane wraps the trivial-cycle $S^3$ inside $S^7$. The trivial 3-cycle implies that there are no different theories classified by it. This embedding is a part of the stable construction in~\cite{Karch2,Arean:2006pk}. Here, the stability of the M5-brane wrapped on $S^3$ is reviewed. Writing the $S^7$ metric as 
 \be
 ds^2=d\xi^2+\cos^2\xi d\Omega^2_{3(1)}+\sin^2\xi d\Omega^2_{3(2)}, 
 \ee
 the M5-brane is wrapped on $S^3_{(1)}$ at $\xi=0$ ($0 \le \xi \le \pi/2$). Expanding around $\xi=0$ and choosing the zero mode of $S^3$ harmonics, the  M5-brane action becomes
 \begin{align}
&S=\sqrt{-G}\Big(1+2R^2_{AdS}G^{ab}\partial_{a}\bar{\xi}\partial_{b}\bar{\xi}-\dfrac{3}{2}\bar{\xi}^2\Big),
 \end{align}
 where $\bar{\xi}$ is the fluctuation of $\xi$.
Then, it is found that $m^2$ of the scalar $\bar{\xi}$ becomes $-3/(4R_{AdS}^2)$ consistent with~\cite{Fiol:2010un}.

\paragraph*{SUMMARY} --- 
In this paper, we studied the $AdS_4\times N(1,1)$/$\mathcal{N}=3$ SCFT correspondence confirmed by the analysis of the moduli space~\cite{Gaiotto:2009tk,Fuj2}. 

We introduced the edge M5-brane filling $AdS_3$ inside $AdS_4$. By analyzing the edge M5-brane, we derived the QHE of the $d=3$ flavored ABJM theory with the C-S levels $(1,-1)$.  We found that the Hall conductivity is quantized such as $\sigma_{xy}=e^2/h$, where the filling fraction is 1.

The same method can also be applied for the gravity dual $AdS_4\times S^7/\mathbb{Z}_k$ of the ABJM theory with C-S levels $(k,-k)$.
Thus, we introduce the M5-brane wrapped on a 3-cycle $S^3/\mathbb{Z}_k$ and can derive the fractionally quantized Hall conductivity $\sigma_{xy}=e^2/kh$. This result was derived in~\cite{Fuj} using the D4-brane in the Type IIA construction. So, we confirmed the M-theory lift. Here, we briefly explain the lift of the Type IIA construction. We consider the D4-brane in the Type IIA theory wrapped on $AdS_3\times \mathbb{CP}^1$ inside $AdS_3\times \mathbb{CP}^3$, where $\mathbb{CP}^1$ is topologically non-trivial. Then, we can construct $S^3/\mathbb{Z}_k$ by making $U(1)$ fibration for the base $\mathbb{CP}^1$. Namely, the M5-brane wrapped on $S^3/\mathbb{Z}_k$ is the M-theory lift of the D4-brane along this $U(1)$ direction. 
The different point from the M5-brane in $AdS_4\times N(1,1)$ is that $S^3/\mathbb{Z}_k$ is the 3-cycle on $S^7/\mathbb{Z}_k$ (for $k\ge 2$). So, the M5-brane wrapped on the 3-cycle $S^3/\mathbb{Z}_k$ is the fractional M5-brane~\cite{Aharony:2008gk}.
 
Then, it will be interesting to discuss the domain walls~\cite{GK} in our setup that changes the rank of one gauge group by 1 (see also the paper~\cite{Ahn}). However, there seems to be only one sphere $S^3$ and the information of 
which rank the M5-brane can change seems to be lost. Moreover, 
 $N(1,1)$ background which can be described by $d=8$ instanton solution contains another KK-monopole that makes the analysis complicated. In the presence of this monopole, we cannot conclude that our edge M5-brane changes the rank of the gauge group on one side of this M5-brane.
 
The M5-brane wrapping on $S^3/\mathbb{Z}_2$ parametrized by  $\alpha=\sigma_1=\sigma_2=\sigma_3=0$ is also interesting. 
As pointed out in~\cite{Hik}, the effect of the  $\mathbb{Z}_2$ Wilson line on $S^3/\mathbb{Z}_2$ should be considered and so using the M5-brane without the manifest $d=6$ covariance seems to be not appropriate. We leave analysis of this M5-brane for future work.

\section{Acknowledgements} 
We would like to thank M. Fukuma, H. Hata, T. Ishii, H. Kawai, T. Takayanagi, and S. Yokoyama for discussions and helpful comments. We would like to thank  Y. Honma and T. S. Tai for valuable discussions and helpful comments. We talked on another M5-brane on $AdS_3\times S^3/\mathbb{Z}_2$ at JPS spring meeting. MF is supported in part by the Japan Society for the Promotion of Science. 



\begin{thebibliography}{99}


\bibitem{Aharony:2008ug}
  O.~Aharony, O.~Bergman, D.~L.~Jafferis and J.~Maldacena,
  JHEP {\bf 0810} (2008) 091

\bibitem{Klebanov:2009sg}
  I.~R.~Klebanov and G.~Torri,
  Int.\ J.\ Mod.\ Phys.\  A {\bf 25}, 332 (2010)
  [arXiv:0909.1580 [hep-th]].

\bibitem{Samtleben:2010eu}
  H.~Samtleben and R.~Wimmer,
  JHEP {\bf 1010}, 080 (2010)
  [arXiv:1008.2739 [hep-th]].

\bibitem{Maldacena:1997re}
  J.~M.~Maldacena,
  Adv.\ Theor.\ Math.\ Phys.\  {\bf 2} (1998) 231
\bibitem{Gub}
  S.~S.~Gubser, I.~R.~Klebanov and A.~M.~Polyakov,
  Phys.\ Lett.\  B {\bf 428}, 105 (1998)
\bibitem{Wit}
  E.~Witten,
  Adv.\ Theor.\ Math.\ Phys.\  {\bf 2}, 253 (1998)

  
 
\bibitem{H}
K. Hosomichi, K. M. Lee, S. Lee, S. Lee and J. Park, 
JHEP 0807, 091 (2008)



\bibitem{Karch}
  A.~Karch and E.~Katz,
  JHEP {\bf 0206}, 043 (2002)

\bibitem{Hohenegger:2009as}
  S.~Hohenegger and I.~Kirsch,
  arXiv:0903.1730 [hep-th].

\bibitem{Gaiotto:2009tk}
  D.~Gaiotto and D.~L.~Jafferis,
  arXiv:0903.2175 [hep-th].

\bibitem{Hik}
  Y.~Hikida, W.~Li and T.~Takayanagi,
  JHEP {\bf 0907}, 065 (2009)
\bibitem{Amm}
  M.~Ammon, J.~Erdmenger, R.~Meyer, A.~O'Bannon and T.~Wrase,
  JHEP {\bf 0911}, 125 (2009)
\bibitem{Jen}
  K.~Jensen,
  Phys.\ Rev.\  D {\bf 82}, 046005 (2010)

  \bibitem{Fuj2}
  M.~Fujita and T.~S.~Tai,
  JHEP {\bf 0909}, 062 (2009)
  
  
\bibitem{GGPT}
  J.~P.~Gauntlett, G.~W.~Gibbons, G.~Papadopoulos and P.~K.~Townsend,
  Nucl.\ Phys.\  B {\bf 500}, 133 (1997)
\bibitem{Baba:2009nv}
  Y.~Baba and T.~S.~Tai,
  JHEP {\bf 1002}, 006 (2010)

\bibitem{Lee:2006ys}
  K.~M.~Lee and H.~U.~Yee,
  JHEP {\bf 0703} (2007) 012
 
\bibitem{Ter1}
  P.~Termonia,
  Nucl.\ Phys.\  B {\bf 577} (2000) 341
\bibitem{Fre1}
  P.~Fre', L.~Gualtieri and P.~Termonia,
  Phys.\ Lett.\  B {\bf 471} (1999) 27


\bibitem{Bil1}
  M.~Billo, D.~Fabbri, P.~Fre, P.~Merlatti and A.~Zaffaroni,
  Class.\ Quant.\ Grav.\  {\bf 18} (2001) 1269
  
\bibitem{Wen:2004ym}
  X.~G.~Wen,
{\it  Oxford, UK: Univ. Pr. (2004) 505 p}

\bibitem{Laughlin:1983fy}
  R.~B.~Laughlin,
  Phys.\ Rev.\ Lett.\  {\bf 50}, 1395 (1983).
 
\bibitem{Kesk}
  E.~Keski-Vakkuri and P.~Kraus,
  [arXiv:0805.4643 [hep-th]].
\bibitem{Davis:2008nv}
  J.~L.~Davis, P.~Kraus and A.~Shah,
  JHEP {\bf 0811}, 020 (2008)
 
\bibitem{Bergman:2010gm}
  O.~Bergman, N.~Jokela, G.~Lifschytz and M.~Lippert,
  arXiv:1003.4965 [hep-th].


\bibitem{Bayntun:2010nx}
  A.~Bayntun, C.~P.~Burgess, B.~P.~Dolan and S.~S.~Lee,
  arXiv:1008.1917 [hep-th].


\bibitem{Fuj}
  M.~Fujita, W.~Li, S.~Ryu and T.~Takayanagi,
  JHEP {\bf 0906}, 066 (2009)
\bibitem{Kawamoto:2009sn}
  S.~Kawamoto and F.~L.~Lin,
  JHEP {\bf 1002}, 059 (2010)
\bibitem{San}
  R.~Santachiara and A.~Tanzini,
  arXiv:1002.5017 [hep-th]
  \bibitem{AGT}
  L.~F.~Alday, D.~Gaiotto and Y.~Tachikawa,
  Lett.\ Math.\ Phys.\  {\bf 91}, 167 (2010) 
\bibitem{Gau}
  J.~P.~Gauntlett, S.~Lee, T.~Mateos and D.~Waldram,
  JHEP {\bf 0508}, 030 (2005)
\bibitem{Pag}
  D.~N.~Page and C.~N.~Pope,
  Phys.\ Lett.\  B {\bf 147}, 55 (1984).


\bibitem{Aharony:2008gk}
  O.~Aharony, O.~Bergman and D.~L.~Jafferis,
  JHEP {\bf 0811}, 043 (2008)
\bibitem{PST}  
P.~Pasti, D.~P.~Sorokin and M.~Tonin,
 Phys.\ Lett.\  B {\bf 398} (1997) 41,\ 
I.~A.~Bandos, K.~Lechner, A.~Nurmagambetov, P.~Pasti, D.~P.~Sorokin and M.~Tonin,
 Phys.\ Rev.\ Lett.\  {\bf 78} (1997) 4332
\bibitem{Aga}
  M.~Aganagic, J.~Park, C.~Popescu and J.~H.~Schwarz,
  Nucl.\ Phys.\  B {\bf 496}, 191 (1997)

\bibitem{Aga2}
  M.~Aganagic, J.~Park, C.~Popescu and J.~H.~Schwarz,
  Nucl.\ Phys.\  B {\bf 496}, 215 (1997)
 
\bibitem{Eli}
  S.~Elitzur, G.~W.~Moore, A.~Schwimmer and N.~Seiberg,
  Nucl.\ Phys.\  B {\bf 326} (1989) 108.

\bibitem{Jensen:2010em}
  K.~Jensen,
  JHEP {\bf 1101}, 109 (2011)
  [arXiv:1012.4831 [hep-th]].



\bibitem{Hun}
  L.~Y.~Hung and A.~Sinha,
  JHEP {\bf 1001}, 114 (2010)
\bibitem{Karch2}
  A.~Karch and L.~Randall,
  JHEP {\bf 0106}, 063 (2001)
  [arXiv:hep-th/0105132].
\bibitem{Arean:2006pk}
  D.~Arean and A.~V.~Ramallo,
  JHEP {\bf 0604}, 037 (2006)
  [arXiv:hep-th/0602174].

\bibitem{Breitenlohner}
  P.~Breitenlohner and D.~Z.~Freedman,
  Phys.\ Lett.\  B {\bf 115}, 197 (1982).

\bibitem{Fiol:2010un}
  B.~Fiol,
  JHEP {\bf 1007}, 046 (2010)
  [arXiv:1005.2133 [hep-th]].




\bibitem{GK}
  S.~S.~Gubser and I.~R.~Klebanov,
  Phys.\ Rev.\  D {\bf 58}, 125025 (1998)
\bibitem{Ahn}
  C.~h.~Ahn,
  Phys.\ Lett.\  B {\bf 466}, 171 (1999)





\end{thebibliography}
\end{document}